\documentclass[conference]{IEEEtran}
\IEEEoverridecommandlockouts
\usepackage{cite}
\usepackage{amsmath,amssymb,amsfonts}
\usepackage{algorithmic}
\usepackage{algorithm}
\usepackage{bm}
\usepackage{bbm}
\usepackage[font=small]{caption}
\usepackage{graphicx}
\usepackage{textcomp}
\usepackage{xcolor}
\usepackage{array}
\usepackage{multirow}
\usepackage{enumitem}
\usepackage{tabulary}
\usepackage{epstopdf}
\usepackage{multicol}
\usepackage{adjustbox}
\usepackage{booktabs}

\usepackage{verbatim}
\usepackage[normalem]{ulem}
\usepackage[subrefformat=parens, labelformat=parens]{subfig}
\def\BibTeX{{\rm B\kern-.05em{\sc i\kern-.025em b}\kern-.08em+T\kern-.1667em\lower.7ex\hbox{E}\kern-.125emX}}
\begin{document}

\title{A Dual-Connection based Handover Scheme for Ultra-Dense Millimeter-Wave Cellular Networks}

\author{
\IEEEauthorblockN{Seongjoon Kang, Siyoung Choi, Goodsol Lee, and Saewoong Bahk}
\IEEEauthorblockA{Department of Electrical and Computer Engineering and INMC, Seoul National University, Seoul, South Korea\\Email: \{sjkang, sychoi, gslee2, sbahk\}@netlab.snu.ac.kr}
}

\maketitle

\begin{abstract}
Mobile users in an ultra-dense millimeter-wave cellular network experience handover events more frequently than in conventional networks, which results in increased service interruption time and performance degradation due to blockages. Multi-connectivity has been proposed to resolve this, and it also extends the coverage of millimeter-wave communications. 
In this paper, we propose a dual-connection based handover scheme for mobile UEs in an environment where they are connected simultaneously with two millimeter-wave cells to overcome frequent handover problems. 
This scheme allows a mobile UE to choose its serving link between the two mmWave connections according to the measured SINRs and then the corresponding base stations may forward duplicate packets to the UE. 
We compare our dual-connection based scheme with a conventional single-connection based scheme through ns-3 simulation. The simulation results show that the proposed scheme significantly reduces handover rate and delay. Therefore, we argue that the dual-connection based scheme helps mobile users achieve performance goals they require in ultra-dense cellular environments.
\end{abstract}

\section{Introduction} 
\label{sec:intro}

The increasing growth in mobile users and traffic demand requires cellular network operators to provide high network capacity \cite{index2015cisco}. As a promising technology in the fifth-generation (5G), millimeter-wave (mmWave) communication provides much higher data rate owing to its new spectrum bands with a  wider bandwidth. In particular, mmWave communication is aggressively pushed ahead by exploiting  28 GHz band first and then higher band \cite{38.901}. However, mmWave signals have a short transmission range and are vulnerable to blockages due to the characteristics of mmWave propagation \cite{rangan2014millimeter}. Specifically, blockages due to buildings, forests, and people, etc. incur Non-line-of-sight (NLOS) connections (i.e., multi-paths) between a mmWave base station (BS) and a mobile user, which greatly lower its signal-to-noise-ratio (SNR) and degrade quality-of-service (QoS) of the user. Hence, LOS connections are highly desirable in mmWave communications.

To overcome blockage effects, multi-connectivity can be exploited where a mobile user is simultaneously connected to multiple BSs, which clearly increases the chance of LOS connections \cite{gupta2018macrodiversity}. However, multi-connections of a user to multiple BSs cause higher handover rate and consequently higher network overhead \cite{bao2016stochastic, zhang2018tractable}. Specifically, the authors in \cite{polese2017improved} proposed a mmWave secondary cell handover (SCH) scheme in an LTE-mmWave dual-connectivity network, which basically uses a single mmWave connection and an LTE connection. 
In this scheme, a mmWave connection is switched to another mmWave cell that provides a larger signal-to-interference-plus-noise-ratio (SINR) after time-to-trigger (TTT). However, even when dynamic TTT is adopted, if the BS density is very high, a TTT based handover scheme increases handover rate to maintain connectivity of the user with better SINR when frequent connection changes occur.

In this paper, we propose a mmWave dual-connectivity scheme that a user equipment (UE) is connected with two mmWave BSs and an LTE BS. By utilizing the two mmWave links in an ultra-dense network, we exploit spatial diversity of the two links, which provides more robust and stable connectivity in a dynamic wireless environment, especially due to the characteristics of mmWave channels. Our proposal helps to decrease handover rate in an ultra-dense cellular network, which contributes to a significant reduction in network control overhead. To support UE mobility under mmWave dual-connectivity, we enhance the SCH scheme by triggering handover according to the channel state between the UE and mmWave BSs.

We compare our dual-connection based SCH scheme with the single-connection based one through ns-3 simulation.
Simulation results show that our scheme outperforms the single-connection based scheme in terms of QoS, network overhead, spectral efficiency, and TCP performance. To the best of our knowledge, we are the first to evaluate the performance of a mobile UE under mmWave dual-connectivity, and propose a handover triggering algorithm suitable for a dynamic mmWave cellular network.

The rest of the paper is organized as follows. Section \ref{sec:relatedWork} summarizes the related work. Section \ref{sec:sysModel} explains the considered network, channel and propagation model. Section \ref{sec:HO4MC} presents our proposed SCH scheme under mmWave dual-connectivity environments. Section \ref{sec:simul} shows simulation results and explains the advantages of the proposed scheme. We conclude the paper in Section \ref{sec:conclu}.

\section{Related Work} 
\label{sec:relatedWork}

An ultra-dense network with very short inter-BS distance and mmWave band operation has been investigated as a promising deployment solution to meet 5G requirements~\cite{ge20165g,baldemair2015ultra}. Multi-connectivity in the ultra-dense mmWave network achieves low session drop probability so that it provides robust communication even when there exist heavy traffic and many blockages \cite{petrov2017dynamic}. The authors in \cite{tatino2018maximum} presented that multi-connectivity in mmWave cellular networks improves overall network throughput compared to single-connectivity.

Through extensive ray tracing simulation using 3D urban building data, they analyzed cell coverage and robustness in mmWave urban cellular network, and revealed the necessity of a multi-frequency heterogeneous mmWave network system \cite{simic2017coverage}.

A mobility handling scheme with multi-connectivity would be quite different from conventional schemes with single- connectivity. Many recent studies  mathematically analyzed handover performance in ultra-dense networks\cite{zhang2018tractable, taufique2018analytical, sadr2015handoff}. In a network with separate control/user plane, the handover probability was investigated in \cite{zhang2018tractable} and the analytic model for handover signalling according to the UE's velocity was proposed in \cite{taufique2018analytical}. In \cite{sadr2015handoff}, the handover rate in a multi-tier heterogeneous network was analyzed.

These works focused only on link level analysis and did not take into account the whole network level performance. The authors in \cite{polese2017improved} conducted several network level simulations using ns-3 and evaluated the performance of the SCH scheme under a single-connection between a mmWave BS and a UE.
However, the single-connection in mmWave communication cannot guarantee reliability and delay constraints because its link quality can be severely degraded due to blockages and UE mobility.

\section{System Model} 
\label{sec:sysModel}

In this section, we describe the system model that we consider. 

\subsection{Network Model}

\begin{figure}[t]
	\centerline{\includegraphics[width=60mm]{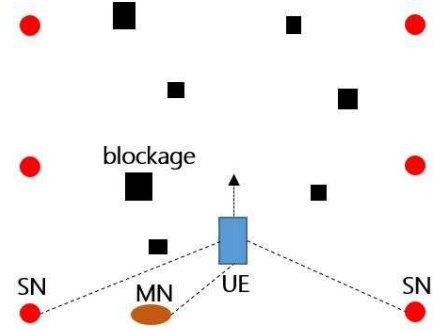}}
	\caption{An illustration of the system model. An MN covers the whole area and SNs are placed at regular intervals. A UE moving at a constant speed has a single-connection to an MN and connections to two SNs simultaneously, and moves along the street among SNs. Blockages are randomly scattered on 2-D plane.}
	\label{fig:networkModel}
\end{figure}

We take account of a downlink scenario where a UE is connected with one master node (MN) and two secondary nodes (SNs) simultaneously. Each BS operates as an MN or SN according to its radio characteristics such as carrier frequency as shown in Fig.~\ref{fig:networkModel}. The MN is responsible for forwarding data traffic from the core network to the two SNs via wired links such as X2 interface and sends control messages toward a UE. On the other hand, the SN works for sending data traffic to a UE with the high data rate.
Hence, we assume that the MN uses the radio access technology (RAT) of low carrier frequency such as LTE where its coverage is much larger than that of an SN.  The SN uses the RAT of the high carrier frequency of mmWave whose data rate is higher than that of the MN.

We assume that the MN covers the whole area we are interested in. SNs are placed at regular intervals, where the distance between the two neighboring SNs depends on the network density. We assume that the two SNs providing dual-connectivity of a UE have different carrier frequencies to avoid inter-cell interference. The UE moves along the path among SNs at a constant speed, maintaining its connectivity with two neighbors SNs. Rectangular-shaped obstacles are  scattered randomly in a 2-dimensional plane to create blockages.

\subsection{Channel and Propagation Model}

We use the mmWave channel model given in \cite{mezzavilla2018end} for a link between an SN and a UE, and the LTE channel model for a link between an MN and a UE. In \cite{mezzavilla2018end}, the authors provide a realistic assessment, including long-term and short-term fading, for mmWave micro- and pico-cellular networks in a densely deployed environment. We assume that mmWave channel matrix, small scale fading and beamforming gain models are the same as  in \cite{giordani2016uplink}.

In this paper, we use the following pathloss model
\begin{equation}
\label{eq:pathloss_logScale}
    PL(d)\,[dB] = \alpha + \beta10\log_{10} (d) + \xi\,, \quad \xi  \sim N (0, \sigma^2)
\end{equation}
where $d$ is the distance between receiver and transmitter, $\sigma^2$ is the lognormal shadowing variance, and $\alpha$ and $\beta$ are the parameters given in \cite{akdeniz2014millimeter}.

We measure channel quality in terms of SINR for serving BS $j$ and UE, which is computed as
\begin{equation}
\label{eq:SINR}
    SINR_{j,UE} = \frac{\frac{P_{j, TX}}{PL_{j,UE}}G_{j,UE}}{\sum_{k\neq j}G_{k,UE}\frac{P_{k,TX}}{PL_{k, UE}} +W_{tot} \times N_{0}},
\end{equation}
where $P_{j, TX}$ is the transmit power of  BS $j$, $G_{j,UE}$ is the beamforming gain, $PL_{j,UE}$ is the pathloss between BS $j$ and UE, $PL_{k, UE}$ is pathloss between BS $k$ and UE,  and $W_{tot} \times N_{0}$ is the thermal noise. SINR values can be classified into three categories: LOS, NLOS, and outage. The classified SINR is used as a criterion for handover decision or path switching in the proposed scheme.

\section{Secondary Cell Handover Scheme for Multi-Connectivity} 
\label{sec:HO4MC}

Multi-connectivity in an ultra-dense network plays an important role in improving the performance of mobile users in terms of throughput, handover rate, and network overhead \cite{zhang2018tractable}. 
We consider mmWave dual-connectivity, where a UE is connected with two mmWave SNs, while keeping a connection to one MN.
One is a serving SN, denoted as $SN_{serve}$, and the other is an idle SN, denoted as $SN_{idle}$. We use only $SN_{serve}$ for data transmission which has higher SINR value enough to meet service requirements, and the other mmWave link plays a role for supporting the serving link intermittently. This leads us to avoid discontinuity of data flow resulting from mmWave channel dynamics and ensure a reliable connection. Swapping the serving link happens when its SINR value is very low.

In our scheme, an MN forwards data traffic from the core network to two SNs, and manages uplink or downlink control signals toward a UE. The two SNs receive a sounding reference signal (SRS) periodically from the UE that they serve, and estimate their downlink channel state under the assumption that channel reciprocity holds \cite{giordani2016uplink}. An MN controls data traffic and makes handover decision according to the channel state information.
\begin{figure}[t]
	\centerline{\includegraphics[width=70mm]{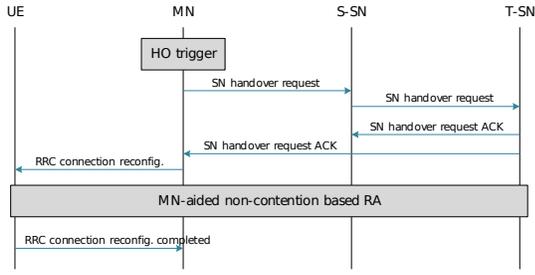}}
	\caption{SN handover procedures.}
	\label{fig:SnHoProcedure}
\end{figure}

There are some benefits of using dual mmWave connections.
First, connecting with the two mmWave SNs enables fast intra-RAT switching so that the MN can choose a mmWave link to serve the UE between the two links without incurring signalling overhead.
 
When the link of $SN_{serve}$ is in NLOS or outage, $SN_{idle}$ serves the UE on behalf of $SN_{serve}$ by switching the data path (not handover) without generating control messages. This reduces the frequency of handover between the SNs and service interruption time. Second, it makes $SN_{serve}$ avoid suffering the buffer overflow problem or data loss in the Radio Link Control (RLC) layer by forwarding RLC data to $SN_{idle}$ via X2 interface. 

In other words, when the SINR value of $SN_{serve}$ is abruptly dropped by blockages, it may incur many packet losses and retransmissions. If $SN_{serve}$ quickly perceives its SINR dropping and forwards its buffered data through X2 interface to $SN_{idle}$ with better link quality, the duration that the link quality of the UE is bad will be reduced, which results in preventing a lot of packet losses. 
Lastly, we apply a packet duplication (PD) scheme for dual-connections, which makes an MN forward the same data to both SNs to exploit diversity gain. On the condition that the two mmWave links have lower SINR values than a certain threshold value or they are unstable, we would get the link diversity gain from the PD scheme.

Since conventional handover schemes are based on single-connectivity between the UE and a mmWave BS, and they do not consider the frequent handover problem in ultra-dense networks. Our dual-connection based handover scheme aims to achieve high reliability and reduce service interruption time.
Dual-connection based handover procedures are shown in Fig.~\ref{fig:SnHoProcedure}, where Radio Resource Control (RRC) reconfiguration messages are exchanged between the UE and MN to exploit wide coverage of the MN unlike in \cite{polese2017improved}.

Assuming that the MN obtains all channel information between the UE and SNs by periodically receiving the link state information from each SN, 
the MN makes all control decisions such as handover, path switching, and PD. Algorithm 1 presents the decision-making procedures at the MN for handover and path switching, which uses the SINR values ($SINR_{serve}$, $SINR_{idle}$) reported by the $SN_{serve}$ and $SN_{idle}$. 

\begin{algorithm}
{\fontsize{9.5pt}{10pt}\selectfont
\caption{Handover decision algorithm}
\label{alg:alg_1}
\begin{algorithmic}[1]
\WHILE{ There is at least one connectable $SN$ }
\IF{ $SINR_{serve}  \leq  SINR_{th}$  \AND  $SINR_{idle} \leq SINR_{th}$ }
        \STATE {Send the same packets to two SNs during HO}
         \IF{ $SINR_{serve}^{target}  \leq SINR_{idle}^{target}$}
            \STATE Handover to $SN_{idle}^{target}$
            \STATE Forward the buffered data to $SN_{idle}^{target}$
         \ELSE
             \STATE Handover to $SN_{serve}^{target}$
        \ENDIF     
\ELSIF{$SINR_{serve} \leq SINR_{th}$ \AND $SINR_{idle} > SINR_{th}$}
    \STATE Switch data forwarding path to $SN_{idle}$
    \STATE Forward the buffered data to $SN_{idle}$
    
\ELSIF{$SINR_{serve}>SINR_{th}$ \AND $SINR_{idle}>SINR_{th}$}
    \IF{ $SINR_{idle} > SINR_{serve}$}
         \STATE Switch data forwarding path to $SN_{idle}$
    \ENDIF
\ENDIF
\ENDWHILE
\end{algorithmic}
}
\end{algorithm}

\begin{itemize}
\item The first case (lines 2 - 9) happens when the SINR values of the two SNs are lower than a certain SINR threshold value, $SINR_{th}$. The MN makes handover decision if there is a target SN that has a greater SINR value than the threshold. The MN knows the SINRs of the two candidates SNs ($SN_{serve}^{target}$, $SN_{idle}^{target}$) from SRS of neighbor SNs. The MN compares the two target SINR values ($SINR_{serve}^{target}$, $SINR_{idle}^{target}$) and selects the higher one. If the selected one is $SN_{idle}^{target}$, $SN_{serve}$ forwards the buffered data to  $SN_{idle}^{target}$ through X2 interface. So only one SN from the two connected ones will be changed. Because the SINR values of the two connected links are expected to be very low, the packet delay would be quite long during the handover event. So, whenever a handover occurs, the MN sends the same data to both SNs via X2 interfaces until all the handover procedures are completed to avoid the long service interruption time.

\item The second case (lines 10 - 12) deals with the case when NLOS or outage occurs in the transmission link and the other idle link has a higher SINR. In this case, the MN knows the channel states and changes the data forwarding path. The MN sends a control message to $SN_{serve}$ to trigger the buffer forwarding to $SN_{idle}$ through X2 interface. 

\item The last case (lines 13 - 16) occurs when the SINR values of the two connected SNs are high enough. The MN selects one SN with a higher SINR and forwards data to it via X2 interface.
\end{itemize}

In the above three cases, handover occurs only when the SINR values of the two connected SNs are smaller than the $SINR_{th}$, which depends on QoS requirements of the UE. Only one mmWave link is used for data transmission except when handover occurs. Provided that the density of SNs is very high, the handover frequency in our scheme would be decreased, compared with that in a single-connection based one.
If the two links are in the outage and there is no candidate link to support handover, the UE falls back to the MN, which is the same as in \cite{polese2017improved}.

Another important consideration in designing the handover scheme is the selection of the TTT value that a UE should wait until handover starts. If the mmWave signal is highly attenuated during the TTT time, too many packet losses occur, resulting in retransmissions. The TTT value affects the frequency of handover a lot in single-connection based schemes, whereas in our scheme it has limited impact on handover rate because handover occurs according to the state of the two links, not a single one. That is, even when the TTT value is very low, unless both the SINR values of the two links are less than $SINR_{th}$, handover does not occur.
\section{Simulation and Analysis}
\label{sec:simul}






\subsection{Simulation Settings and Scenario}

\begin{figure}[t]
	\centerline{\includegraphics[width=1.0\columnwidth]{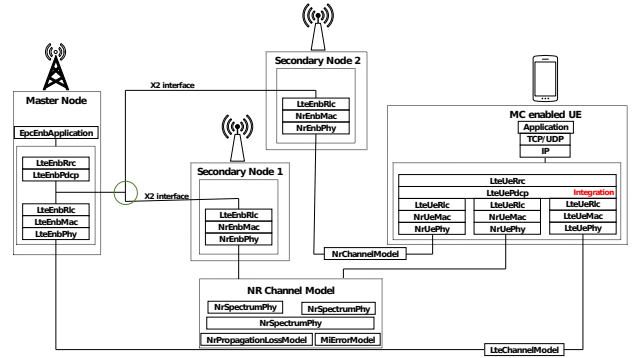}}
	\caption{User-plane protocol model in ns-3 simulator supporting dual-connectivity.}
	\label{fig:dcDLProtocol}
\end{figure}

We modified the uplink based initial access scheme (IA) of NYU simulator \cite{mezzavilla2018end} to make a UE simultaneous connections with one MN and two SNs. The two SNs connected  with the UE use different mmWave carrier frequencies to avoid inter-cell interference. While the UE is moving with dual connections, it periodically tries to find its optimal beam direction toward SNs. To this end, we developed a beam alignment scheme suitable for dual connections. Fig.~\ref{fig:dcDLProtocol} shows the user-plane protocol model in ns-3 that supports dual connections. We added all the control procedures related to handover and data path switching into RRC classes in MN and UE. The network device of the UE is equipped with the protocol stacks corresponding to two SNs and one MN, and all data traffic received from each RLC layer is aggregated at the Packet Data Convergence Protocol (PDCP) class in the UE.

The parameters we used in the simulation are based on realistic system design and summarized in Table \ref{tab:table1}. Low carrier frequency is employed at the MN while ultra high carrier frequency at SNs because the MN's coverage is much larger than that of an SN, ensuring stable signalling message exchange with the UE. Specifically, LTE and mmWave protocols are installed at the MN and SNs, respectively. 
We consider UDP and RLC AM protocols to remove unnecessary delay and to ensure transmission reliability. Considering the coverage of a mmWave BS \cite{simic2017coverage}, we emulate the deployment of mmWave BSs in a dense urban region, where six BSs are located in an area of $100\times100$~m$^2$, setting the inter-BS distance as 50~m. The UE moves along the street between mmWave BSs at a constant speed of 10~m/s. Small blockages indicating humans, trees, vehicles, etc., are randomly scattered to make mmWave channels more dynamic. They have rectangular shapes with their widths uniformly distributed between 1.0 and 2.0~m.

\begin{table}[!ht]
  \begin{center}
    \caption{Simulation parameters}
    \label{tab:table1}
    \begin{tabular}{|c|c|} 
    \hline
      \textbf{Parameter} & \textbf{Value}\\
      \hline
      mmWave BS TX power (dBm) & 30\\
      mmWave bandwidth (MHz) & 1000 \\
      mmWave BS antenna configuration & $8\times8$ ULA\\
      UE antenna configuration & $4\times4$ ULA\\
      Inter-BS distance (m) & 50 \\
      RLC mode & RLC AM \\
      RLC buffer size (MB) & 100\\
      LTE bandwidth (MHz) & 20 \\
      UE speed (m/s) & 10 \\
      LTE downlink carrier frequency (GHz) & 2.1\\
      X2 link delay (ms) & 1.0 \\
      file size for downloading (MB) & 0.1, 1, 10, 100, 200\\
      file transmission interval (ms) & 120 \\
      blockage size (m) & x, y dimensions $\in ( 1.0,~2.0 )$ \\
      blockage density (blockages/km$^2$) & 1000, 2000, 4000, 6000 \\
      TTT (ms) & 20 \\
      $SINR_{th}$ (dB) & 20 \\
      \hline
    \end{tabular}
  \end{center}
\end{table}

For comparison, we consider the single-connection based scheme that has connections to one MN and one SN in \cite{polese2017improved}.
To evaluate the benefits of using dual-connection over single-connection, we simulate a file download scenario and use performance metrics of file download completion time, handover rate, file download failure ratio, and TCP throughput.
In the simulation, the serving SN periodically sends a fixed sized file to the UE with the interval of 120~ms. 



\subsection{Simulation Results and Discussions}
\begin{figure}[t]
	\centerline{\includegraphics[width=70mm]{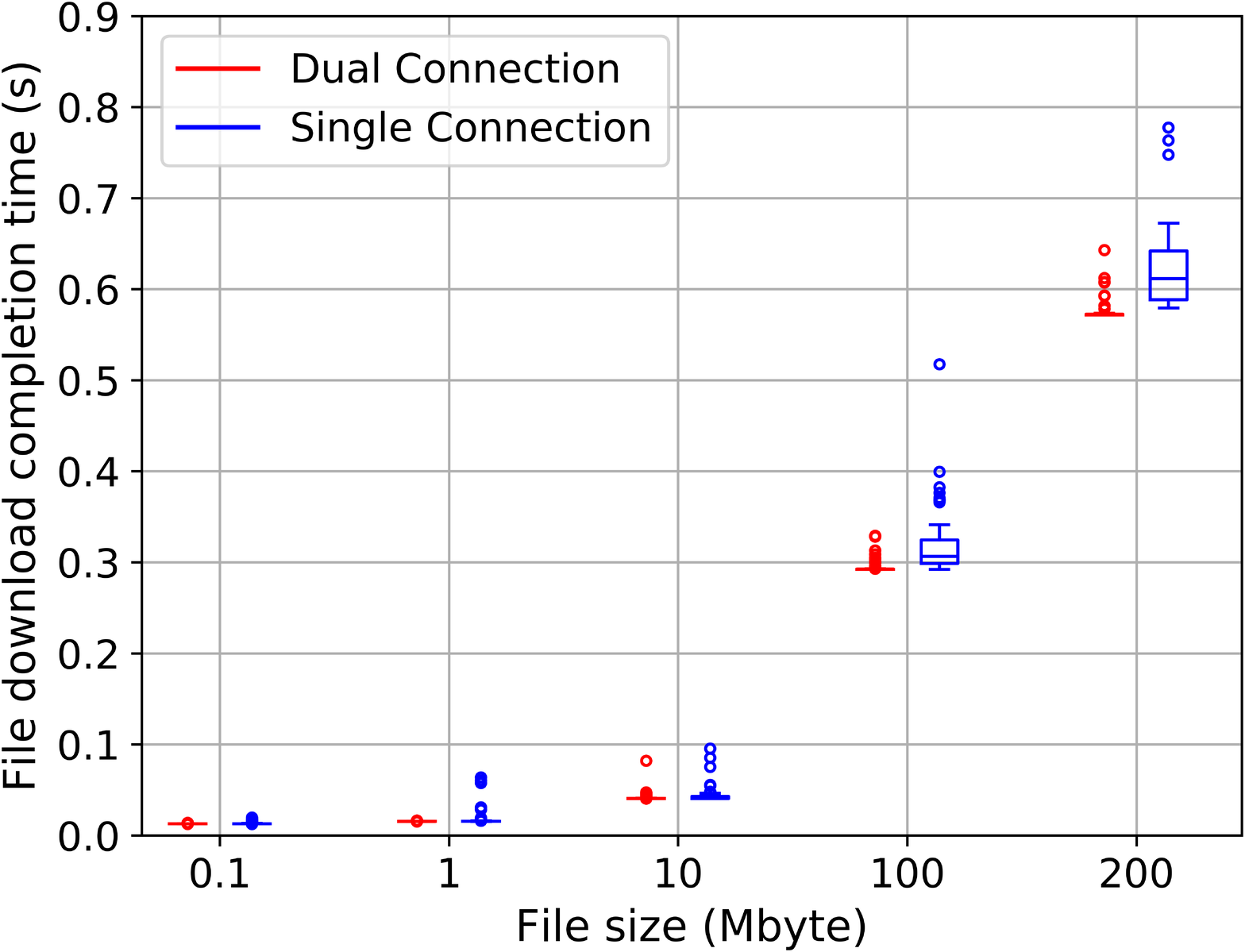}}
	\caption{File download completion time with 4,000 blockages/km$^2$.}
	\label{fig:fileDownloadTime}
\end{figure}

We measured file download completion time according to the file size as shown in Fig.~\ref{fig:fileDownloadTime}. We observe that the dual-connection based scheme achieves smaller download completion time by reducing the number of outlier points, which can be interpreted as service interruption events. From these results, we noticed that utilizing an additional mmWave link improves QoS performance. In particular, when the file size is very large as much as 100 or 200~Mbytes, the file download time using single-connection is much longer than that of using dual-connection, which means that large file transmission on single-connection is unlikely to guarantee QoS. In the dual-connection based scheme, there are a few service interruption events. The reason is that channel state information reported from SNs may not always help to estimate link state perfectly. Note that received SINR values may change a lot after SRS reporting since the wireless environment is dynamic.

\begin{figure}[t]
	\centerline{\includegraphics[width=70mm]{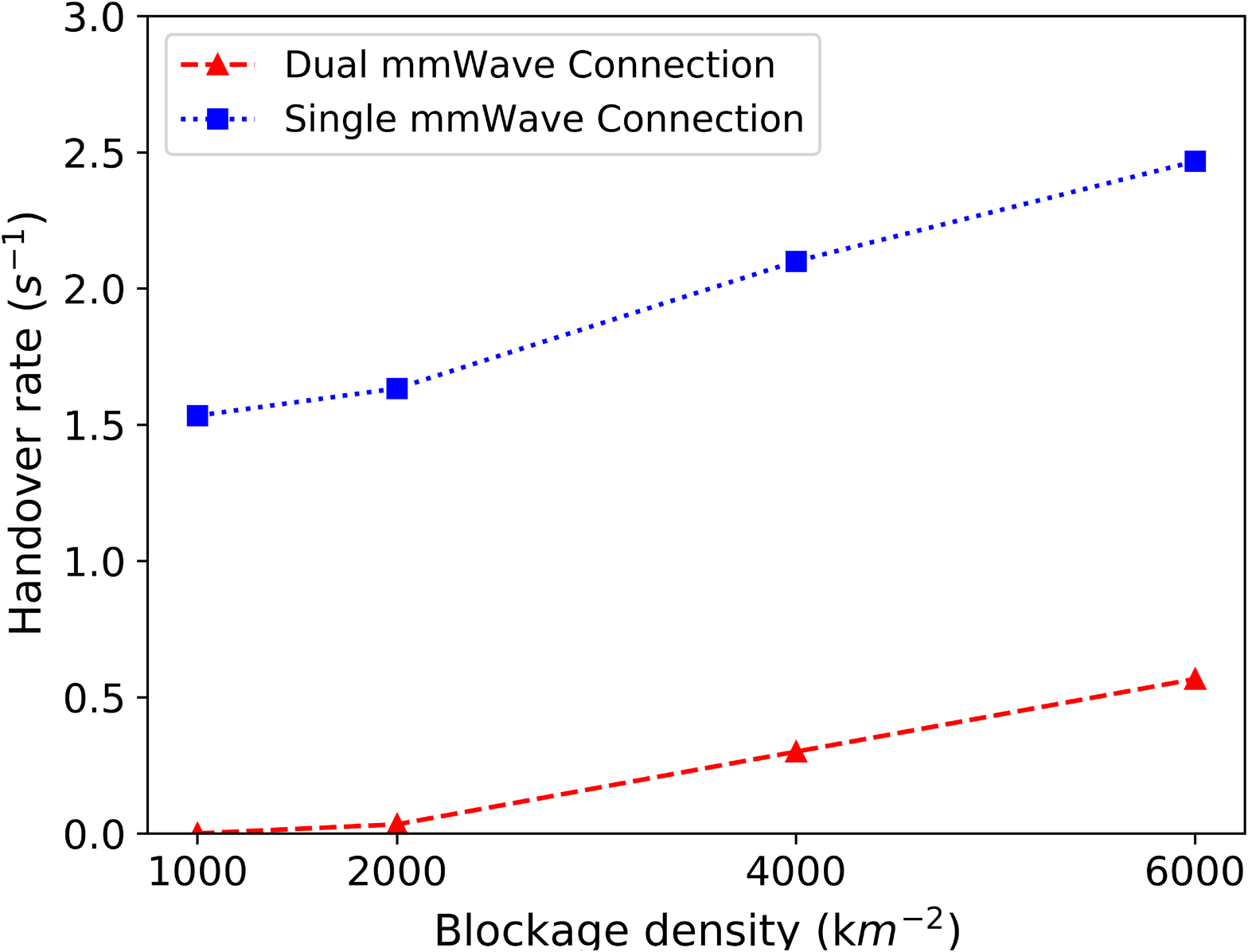}}
	\caption{Handover rate with the transmission file size of 1 Mbyte.}
	\label{fig:handoverRate}
\end{figure}

We evaluate handover rate and transmission reliability under different blockage densities. We define the handover rate as the total number of handover trials divided by the whole simulation time. Although the BS density is high, under high blockage density, it would be difficult to maintain a stable connection between the UE and BS due to frequent handovers, resulting in falling back to the MN. If we have an extra link, we can exploit spatial diversity and reduce control signalling overhead caused by handover or falling back. Fig.~\ref{fig:handoverRate} shows the handover rate for single- and dual-connection. Our dual-connection based scheme remarkably reduces the number of handovers and accordingly overall network overhead. We observe that the UE does not experience handover when the blockage density is below 1000 blockages/km$^2$, i.e., very low. In this case, switching the data path between the two connected links is enough.

\begin{figure}[t]
	\centerline{\includegraphics[width=70mm]{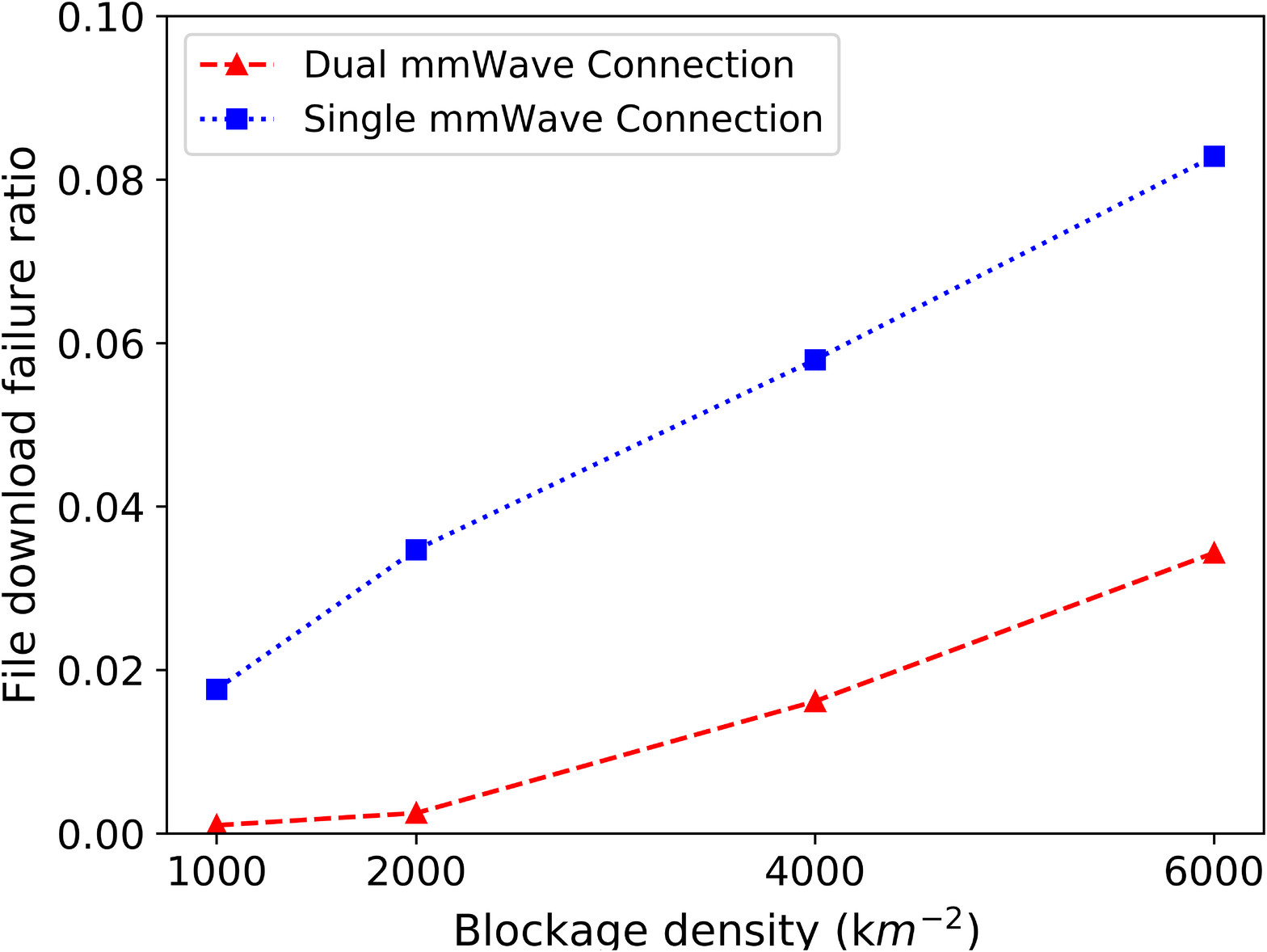}}
	\caption{File download failure ratio.}
	\label{fig:fileDownloadFailure}
\end{figure}
\begin{figure}[t]
    \centering
\mbox{\subfloat[CWND variation]{\label{subfig:tcp_cwnd}
      \includegraphics[width=.70\columnwidth, trim = 0 0 0 0]{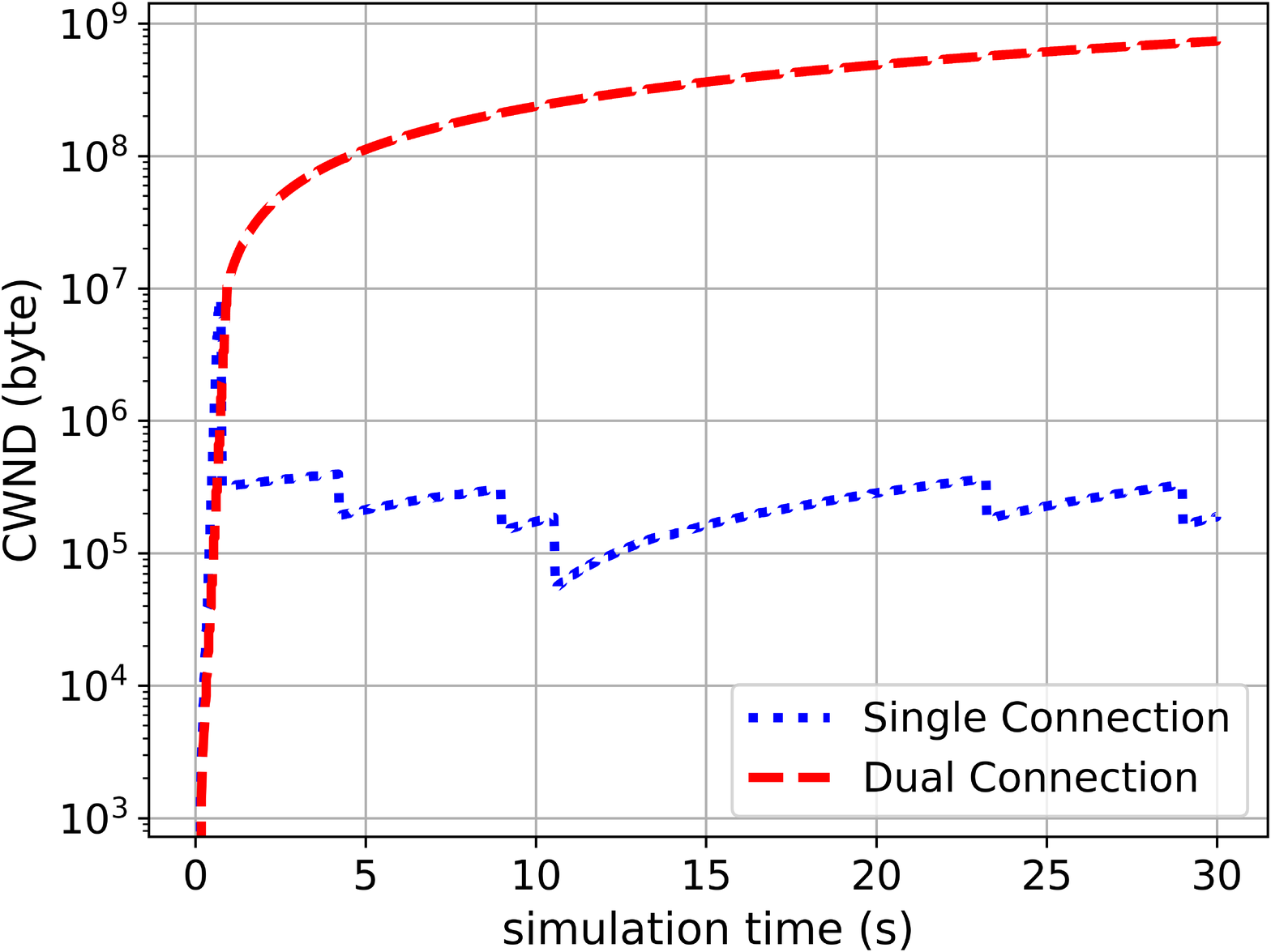}}}
      \quad
\mbox{\subfloat[Throughput]{\label{subfig:tcp_tput}
      \includegraphics[width=.70\columnwidth, trim = 0 0 0 0]{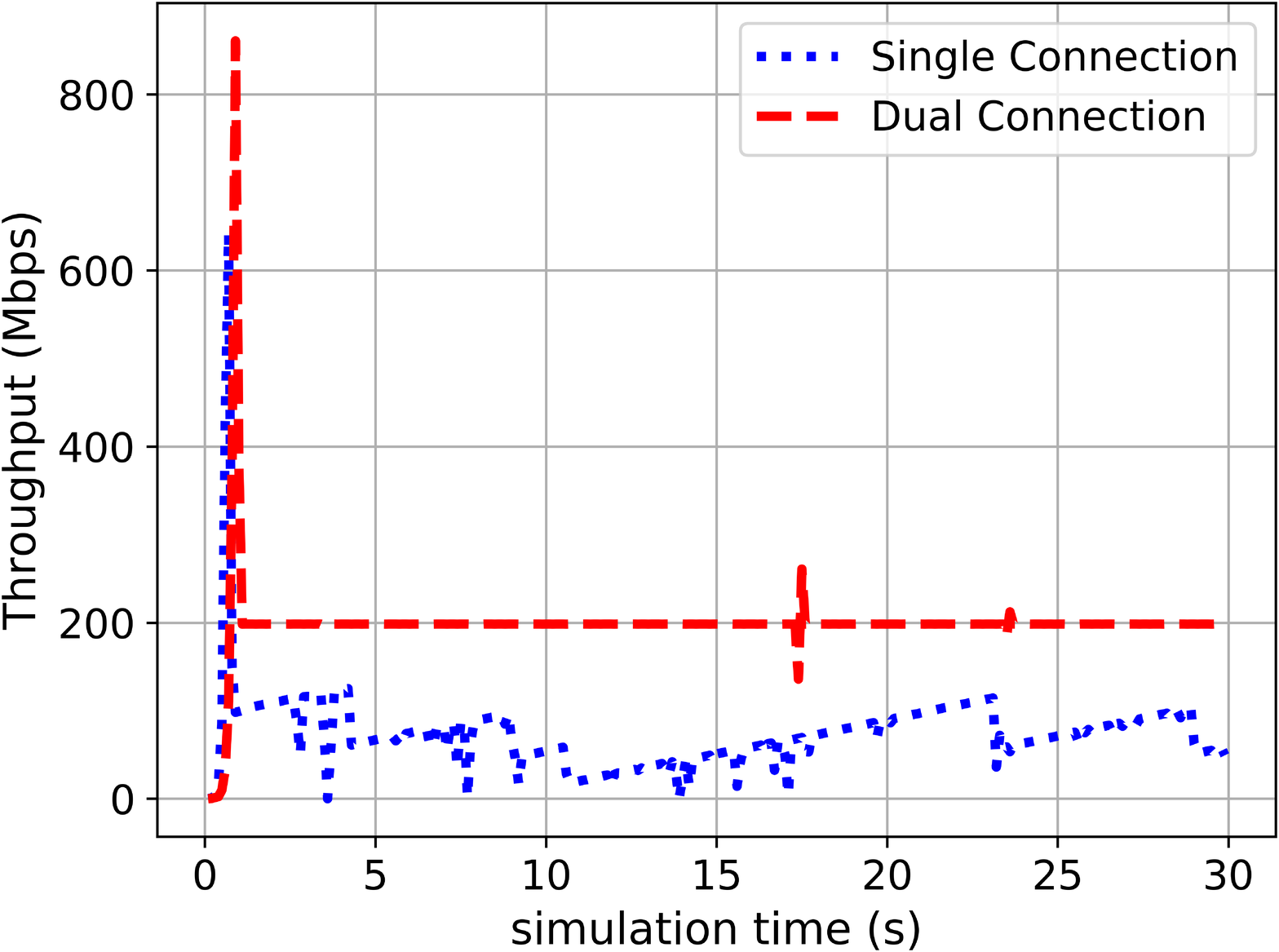}}}
\caption{TCP performance comparison}
\label{fig:tcp_performance}
\end{figure}


To see transmission reliability, we take into account the file download failure ratio, which is defined as the number of file download failures divided by that of total file transmissions. A delay constraint is given to assess whether a file is successfully downloaded or not. If the file does not arrive within the delay constraint, it is regarded as a transmission failure. Fig.~\ref{fig:fileDownloadFailure} shows the file download failure ratio when blockage density varies, and indicates that the dual-connection scheme achieves reliable file transmission. However, like the single connection scheme, the dual-connection scheme doesn't guarantee high reliability either when the blockage density is high. This is because, even if we utilize two mmWave connections, under high blockage density, handover or falling back events occur frequently.

Lastly, we assess TCP (NewReno) performance in the same scenario and simulation settings; the UE moves up and down with constant speed under the blockage density of 4000 blockages/km$^2$. A remote server sends data to the UE with rate 200 Mbps during the simulation time, and the PDCP reordering function is enabled for aggregating packets from the two paths. Figs.~\subref*{subfig:tcp_cwnd} and~\subref*{subfig:tcp_tput} show that our dual-connection based scheme outperforms the single-connection based one in terms of TCP throughput and CWND (congestion window) variation. This is because the proposed scheme successfully reduces both service interruption time and packet losses caused by frequent handover events.

\section{Conclusion}
\label{sec:conclu}
In this paper, we proposed a handover scheme that considers dual-connectivity in the mmWave band, and evaluated its performance through ns-3 simulation. For doing this, we modified the mmWave ns-3 modules that have been implemented by NYU. Our handover scheme considers the case that a UE is connected to two mmWave BSs and one LTE BS. In the  simulation, we compared the performance of the dual-connection based scheme with that of the single-connection based one, in terms of file download time, hand over rate, file download failure ratio, and TCP throughput. 

When it comes to the use of radio resource, the dual-connection based scheme utilizes twice as many controlling signals as a single-connection based scheme to keep constant connections with a UE. However, our scheme achieves a more robust and stable performance compared to conventional single-connection based handover scheme although there are overheads of keeping the dual-connection with a UE.

Our scheme shows more reduced and less fluctuating download completion time according to the file size, compared to the single connection based scheme. By varying the blockage density, we observed the handover rate and download failure ratio. Our scheme shows significantly lowered handover rate and download failure ratio, resulting in much better reliability.
Lastly, we observed our handover scheme achieves very much improved TCP performance in terms of throughput and CWND variation.

\section*{Acknowledgment}
\thanks{This research was supported by Basic Science Research Program through the National Research Foundation of Korea~(NRF) funded by the Ministry of Science, ICT and Future Planning (No. 2017R1E1A1A01074358)}
\bibliographystyle{IEEEtran}
\scriptsize
\bibliography{main.bbl}

\begin{thebibliography}{10}
\providecommand{\url}[1]{#1}
\csname url@samestyle\endcsname
\providecommand{\newblock}{\relax}
\providecommand{\bibinfo}[2]{#2}
\providecommand{\BIBentrySTDinterwordspacing}{\spaceskip=0pt\relax}
\providecommand{\BIBentryALTinterwordstretchfactor}{4}
\providecommand{\BIBentryALTinterwordspacing}{\spaceskip=\fontdimen2\font plus
\BIBentryALTinterwordstretchfactor\fontdimen3\font minus
  \fontdimen4\font\relax}
\providecommand{\BIBforeignlanguage}[2]{{%
\expandafter\ifx\csname l@#1\endcsname\relax
\typeout{** WARNING: IEEEtran.bst: No hyphenation pattern has been}%
\typeout{** loaded for the language `#1'. Using the pattern for}%
\typeout{** the default language instead.}%
\else
\language=\csname l@#1\endcsname
\fi
#2}}
\providecommand{\BIBdecl}{\relax}
\BIBdecl

\bibitem{index2015cisco}
C.~V.~N. Index, ``{Cisco visual networking index: global mobile data traffic
  forecast update, 2014--2019},'' \emph{Tech. Rep}, 2015.

\bibitem{38.901}
{3GPP TR 38.901}, ``{Study on channel model for frequencies from 0.5 to
  100~GHz},'' ver. 14.0.0, Dec. 2017.

\bibitem{rangan2014millimeter}
S.~Rangan, T.~S. Rappaport, and E.~Erkip, ``{Millimeter-wave cellular wireless
  networks: Potentials and challenges},'' \emph{Proc. IEEE}, vol. 102, no.~3,
  pp. 366--385, 2014.

\bibitem{gupta2018macrodiversity}
A.~K. Gupta, J.~G. Andrews, and R.~W. Heath, ``{Macrodiversity in cellular
  networks with random blockages},'' \emph{IEEE Trans. Wireless Commun.},
  vol.~17, no.~2, pp. 996--1010, 2018.

\bibitem{bao2016stochastic}
W.~Bao and B.~Liang, ``{Stochastic geometric analysis of handoffs in
  user-centric cooperative wireless networks.}'' in \emph{Proc. IEEE INFOCOM},
  Apr. 2016, pp. 1--9.

\bibitem{zhang2018tractable}
H.~Zhang and W.~Huang, ``{Tractable Mobility Model for Multi-Connectivity in 5G
  User-Centric Ultra-Dense Networks},'' \emph{IEEE Access}, vol.~6, pp.
  43\,100--43\,112, 2018.

\bibitem{polese2017improved}
M.~Polese, M.~Giordani, M.~Mezzavilla, S.~Rangan, and M.~Zorzi, ``{Improved
  handover through dual connectivity in 5G mmWave mobile networks},''
  \emph{IEEE J. Sel. Areas Commun.}, vol.~35, no.~9, pp. 2069--2084, 2017.

\bibitem{ge20165g}
X.~Ge, S.~Tu, G.~Mao, C.-X. Wang, and T.~Han, ``{5G ultra-dense cellular
  networks},'' \emph{IEEE Wireless Commun.}, vol.~23, no.~1, pp. 72--79, 2016.

\bibitem{baldemair2015ultra}
R.~Baldemair, T.~Irnich, K.~Balachandran, E.~Dahlman, G.~Mildh, Y.~Sel{\'e}n,
  S.~Parkvall, M.~Meyer, and A.~Osseiran, ``{Ultra-dense networks in
  millimeter-wave frequencies},'' \emph{IEEE Commun. Mag.}, vol.~53, no.~1, pp.
  202--208, 2015.

\bibitem{petrov2017dynamic}
V.~Petrov \emph{et~al.}, ``{Dynamic multi-connectivity performance in
  ultra-dense urban mmWave deployments},'' \emph{IEEE J. Sel. Areas Commun.},
  vol.~35, no.~9, pp. 2038--2055, 2017.

\bibitem{tatino2018maximum}
C.~Tatino, I.~Malanchini, N.~Pappas, and D.~Yuan, ``{Maximum throughput
  scheduling for multi-connectivity in millimeter-wave networks},'' in
  \emph{Proc. IEEE WiOpt}, May 2018, pp. 1--6.

\bibitem{simic2017coverage}
L.~Simic, S.~Panda, J.~Riihijarvi, and P.~Mahonen, ``{Coverage and robustness
  of mm-wave urban cellular networks: multi-frequency HetNets are the 5G
  future},'' in \emph{Proc. IEEE SECON}, Jun. 2017, pp. 1--9.

\bibitem{taufique2018analytical}
A.~Taufique, A.~Mohamed, H.~Farooq, A.~Imran, and R.~Tafazolli, ``{Analytical
  Modelling for Mobility Signalling in Ultra-Dense HetNets},'' \emph{IEEE
  Trans. Veh. Technol.}, 2018.

\bibitem{sadr2015handoff}
S.~Sadr and R.~S. Adve, ``{Handoff rate and coverage analysis in multi-tier
  heterogeneous networks},'' \emph{IEEE Trans. Wireless Commun.}, vol.~14,
  no.~5, pp. 2626--2638, 2015.

\bibitem{mezzavilla2018end}
M.~Mezzavilla, M.~Zhang, M.~Polese, R.~Ford, S.~Dutta, S.~Rangan, and M.~Zorzi,
  ``{End-to-end simulation of 5{G} mmwave networks},'' \emph{IEEE Commun.
  Surveys Tuts.}, 2018.

\bibitem{giordani2016uplink}
M.~Giordani, M.~Mezzavilla, S.~Rangan, and M.~Zorzi, ``{Uplink-based framework
  for control plane applications in 5G mmWave cellular networks},'' \emph{IEEE
  Trans. Wireless Commun.}, {submitted. [Online]. Available:
  \url{http://arxiv.org/abs/1610.04836}}.

\bibitem{akdeniz2014millimeter}
M.~R. Akdeniz, Y.~Liu, M.~K. Samimi, S.~Sun, S.~Rangan, T.~S. Rappaport, and
  E.~Erkip, ``{Millimeter wave channel modeling and cellular capacity
  evaluation},'' \emph{IEEE J. Sel. Areas Commun.}, vol.~32, no.~6, pp.
  1164--1179, 2014.

\end{thebibliography}

\end{document}